\documentclass[12pt]{article}

\usepackage{newtxtext,newtxmath}

\usepackage{hyperref}
\usepackage{pdflscape}

\usepackage{graphicx}

\usepackage[letterpaper,margin=1in]{geometry}

\renewenvironment{abstract}
	{\quotation}
	{\endquotation}

\date{}

\makeatletter
\renewcommand{\fnum@figure}{\textbf{Figure \thefigure}}
\renewcommand{\fnum@table}{\textbf{Table \thetable}}
\makeatother

\newcommand{\qcode}[3]{$[\![ #1, #2, #3 ]\!]$}

\usepackage{scicite}
\usepackage{url}

\def\scititle{
    Efficient and Universal Neural-Network Decoder for Stabilizer-Based Quantum Error Correction
}
\title{\bfseries \boldmath \scititle}

\author{
	Gengyuan Hu$^\text{1}$,
        Wanli Ouyang$^{\text{1,2}\ast}$,
        Chao-Yang Lu$^\text{3,4,5,6}$,
	Chen Lin$^{7\ast}$,
        Han-Sen Zhong$^{\text{1,8}\ast}$\and
        \small$^{1}$Shanghai Artificial Intelligence Laboratory, Shanghai \& 200030, China.\and
        \small$^{2}$Department of Information Engineering, The Chinese University of Hong Kong, Hong Kong SAR \& HKG, China.\and
        \small$^{3}$Hefei National Research Center for Physical Sciences at the Microscale and School of \and
        \small Physical Sciences, University of Science and Technology of China, Hefei \& 230026, China.\and
        \small$^{4}$CAS Center for Excellence in Quantum Information and Quantum Physics, University of \and
        \small Science and Technology of China, Shanghai \& 201315, China.\and
        \small$^{5}$Hefei National Laboratory, University of Science and Technology of China, Hefei \& 230088, China.\and
        \small$^{6}$Shanghai Research Center for Quantum Sciences, Shanghai \& 201315, China.\and
        \small$^{7}$Department of Engineering, University of Oxford, Oxford \& OX1 4BH, UK.\and
        \small$^{8}$Shanghai Innovation Institute, Shanghai \& 200231, China.\and
	\small$^\ast$Corresponding author. Email: wlouyang@ie.cuhk.edu.hk \and
	\small$^\ast$Corresponding author. Email: chen.lin@eng.ox.ac.uk \and
	\small$^\ast$Corresponding author. Email: zhonghansen@pjlab.org.cn \and
}

\begin{document} 

\maketitle
\clearpage
\begin{abstract} \bfseries \boldmath

Scaling quantum computing to practical applications necessitates reliable quantum error correction. Despite the proposal of numerous correction codes, the overall correction efficiency remains critically limited by decoding algorithms. We introduce GraphQEC, a code-agnostic decoder leveraging machine-learning on the graph structure of stabilizer codes with linear time complexity. GraphQEC demonstrates unprecedented accuracy and efficiency across all tested code families, including surface codes, color codes, and quantum low-density parity-check (QLDPC) codes. 
For instance, under physical error rates of p = 0.005, GraphQEC achieves a logical error rate of $9.55 \times 10^{-5}$ on a distance-12 QLDPC code. This represents an 18-fold improvement over the best specialized decoder ($1.74 \times 10^{-3}$) while maintaining a decoding speed of $157\mu$s/cycle.
Our approach represents a significant step toward universal real-time quantum error correction across diverse stabilizer codes.

\end{abstract}

\noindent
\section*{Main}
Quantum computing represents a revolutionary computational paradigm with the potential to transform fields ranging from cryptography and materials science to drug discovery and artificial intelligence\cite{Shor1994,Grover1996,HHL2009,VQE2022}. Recent experimental demonstrations have showcased quantum advantage across various computational tasks\cite{arute2019quantum,zhong2020quantum,google2021exponential,wu2021strong,ai2024quantum,gao2025establishing}, marking significant progress toward practical applications. However, realizing the full potential of quantum computing requires overcoming the fundamental challenge of decoherence and gate errors. Quantum error correction (QEC) addresses this challenge by encoding logical qubits across multiple physical qubits, enabling fault-tolerant quantum computation even with imperfect hardware components\cite{kitaev2003fault,Fowler2012Surface,gottesman2013fault}.

QEC comprises two essential processes: encoding quantum information into protected logical states and decoding syndrome measurement results to identify and correct errors. The stabilizer formalism provides a mathematical framework for constructing quantum error correction codes, where a set of commuting multi-qubit Pauli operators (stabilizers) define a codespace\cite{Gottesman1997Stabilizer,Nielsen2010Quantum}.

Measurements of these stabilizer operators provide error syndromes and projecting the system state onto eigenspaces of the stabilizers, which is also the codespace for the logical qubits, allowing error detection without collapsing the encoded quantum information.
The structure of stabilizer codes inherently forms a graph representation, where qubits correspond to nodes and stabilizer measurements establish relationships between them. This graph structure defines the code's error detection capabilities and correction properties. The measurement outcomes of stabilizer generators remain invariant in the absence of errors, with deviations in sequential measurements indicating the occurrence of errors in the system.


Effective quantum error decoders offer two quantifiable advantages. First, decoder time complexity directly impacts the feasibility of real-time error correction, which is necessary for maintaining quantum coherence during computation. Linear-time decoders are particularly valuable for extended quantum circuits requiring multiple error correction cycles. Second, decoder accuracy directly influences the logical error rate, which determines the threshold behavior and resource requirements for fault-tolerant computation\cite{Wang2003confinement}. Improved decoder accuracy effectively increases the threshold value, reducing the physical qubit overhead required to achieve a target logical error rate.

Current decoder implementations face significant technical limitations. Matching-based decoders, including minimum-weight perfect matching algorithms, demonstrate efficient performance for surface codes but are fundamentally limited to pairwise error correlations, making them unsuitable for codes with higher-order correlations or complex measurement noise patterns\cite{Dennis2002MWPM,higgott2023sparse}.
Belief propagation (BP) decoders frequently encounter convergence failures when applied to quantum codes due to the presence of short cycles in their factor graphs, necessitating computationally expensive post-processing techniques such as ordered statistics decoding (OSD)\cite{roffe_decoding_2020,Roffe_LDPC_Python_tools_2022}. Machine learning decoders have emerged as a promising approach in recent years\cite{torlai2017neural,baireuther2019neural,maskara2019advantages,varbanov2023neural,cao2023qecgpt,bausch2024learning}. However, the previous work is either optimized only for specific quantum error correction codes or fails to achieve comparable performance to traditional decoders when applied to highly complex codes and circuit-level noise.

Here we present GraphQEC, a temporal graph neural network (GNN) decoder that addresses these limitations. By operating directly on stabilizer code graph structures\cite{zhou2018graph}, GraphQEC requires no code-specific design modifications, enabling scalable quantum error correction across diverse code families and noise conditions.

To demonstrate the universality of our approach, we conducted comprehensive circuit-level noise simulations on three distinct stabilizer code families: color codes\cite{bombin2006topological}, Bivariate Bicycle (BB) codes\cite{bravyi2024high,wang2025demonstration}, and surface codes\cite{kitaev2003fault,Bon2021XZZX}. Our results show substantial improvements in logical error rates compared to existing decoders across all code families under realistic noise conditions. Critically, these performance gains are achieved without sacrificing computational efficiency, as the decoder maintains linear time complexity, which is essential for practical deployment as quantum algorithms require many sequential error correction rounds without accumulating computational delays.

These results establish GraphQEC as a practical universal decoder for quantum error correction, combining superior accuracy with computational efficiency across diverse code families. By providing a single, high-performance solution for any stabilizer code, our work simplifies the path to fault-tolerant quantum computing and enables rapid evaluation of emerging quantum error correction schemes.


\subsection*{Decode on the Tanner graph}

Analogously to classical digital circuits, a common way to represent quantum logical operations is using the quantum circuit model. In the quantum circuit model, quantum computations are executed through sequences of elementary quantum gates acting on qubits. However, physical implementations are susceptible to errors during all operations, including gate operations, qubit initialization, and measurements. This realistic scenario can be characterized by the circuit-level noise model.

To protect quantum information against such noise, stabilizer codes are employed. A \qcode{n}{k}{d} stabilizer code is defined by a set of $n-k$ stabilizer generators, represented by a $2n\times(n-k)$ binary symplectic matrix $H \equiv \left[\cdot|\cdot\right]$. Each row in $H$ represents a stabilizer generator, and each column of each half corresponds to a qubit. The matrix elements specify Pauli operators through the mapping:
\begin{equation}\label{eq:sympletic representation}
    (0|0) \rightarrow I, (0|1) \rightarrow Z, (1|0) \rightarrow X, (1|1) \rightarrow Y
\end{equation}
such that each stabilizer generator is a product of Pauli operators on the encoded qubits.

\begin{figure}
\centering
\includegraphics[width=\textwidth]{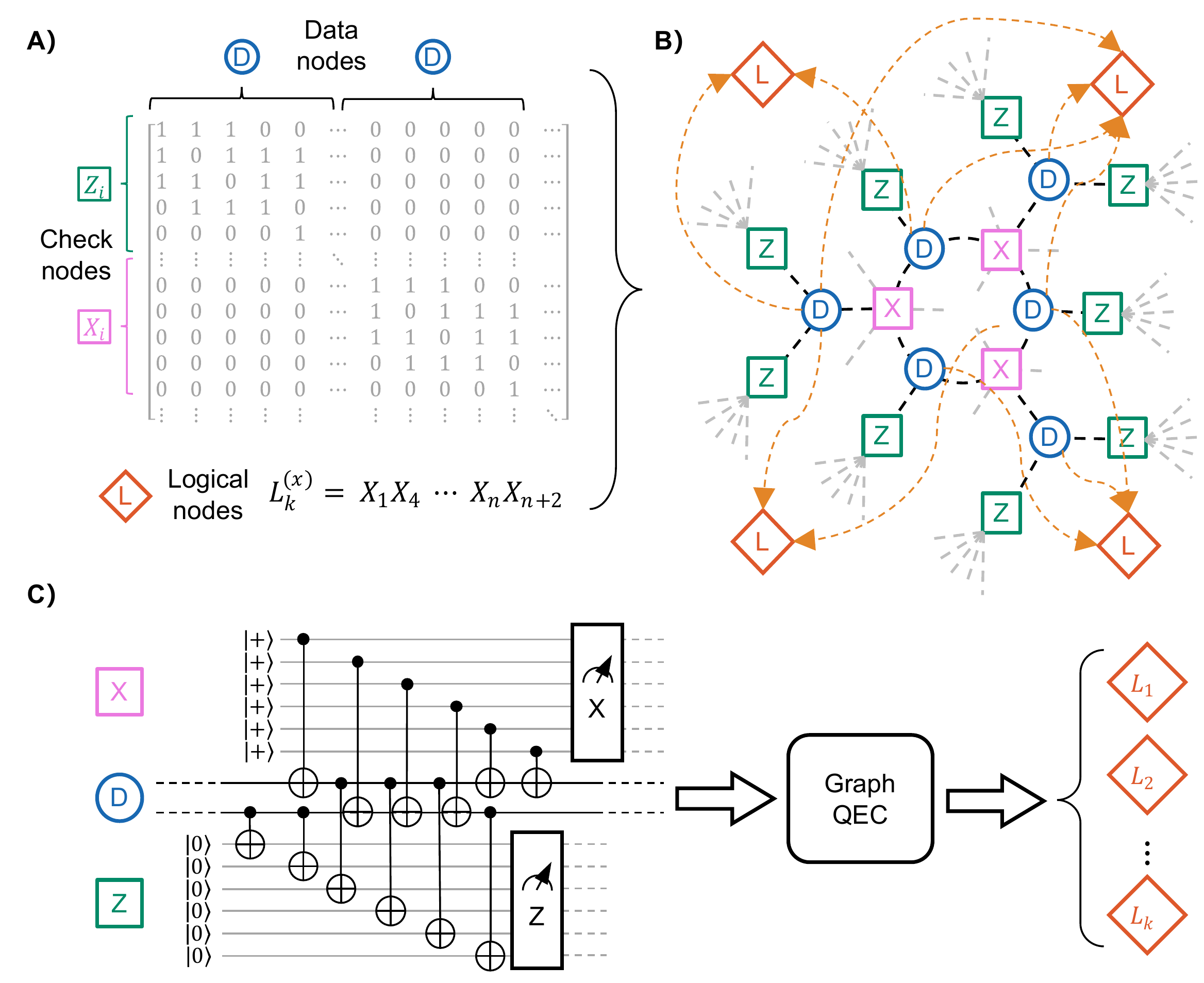}
\caption{\textbf{Overview of the stabilizer code decoding framework}. (A) The stabilizer code can be represented by a stabilizer generator matrix, and the logical qubit is defined by a set of selected single-qubit pauli operators. (B) The stabilizer code can be equivalently represented by an extended Tanner Graph, with physical qubits become data nodes, stabilizer generators become check nodes, and logical pauli operators become logical nodes. (C) The syndrome measurement circuit generate the syndromes repeatedly. The GraphQEC decoder receives the syndromes and decode it on the extended Tanner graph.
}\label{WorkingFlow}
\end{figure}

Our approach begins by representing this stabilizer structure as an extended Tanner graph (Fig.~\ref{WorkingFlow}). In this representation, stabilizer generators become check nodes and qubits become data nodes, with edges indicating non-zero matrix elements. The Tanner graph serves not only as a mathematical equivalent representation for a stabilizer code but also as an abstracted description of the code's hardware topology. Signals from the quantum device implementing the described code can be naturally embedded into the graph for feature extraction. Unlike the decoding graph used in traditional decoders like MWPM and BP-OSD, where the decoding graph is also called the Tanner graph in their context, the Tanner graph here is fully agnostic to the circuit noise, we focus on the mathematical structure of a stabilizer code, leaving the circuit noise details to be learned in the training process.

During quantum error correction, measurements from each stabilizer generator are collected in successive cycles until the logical qubit measurement is required. The collected measurement results from previous cycles, called syndromes, are used to correct the logical measurement results through decoding. To connect the decoding problem with the syndrome, we extend the basic Tanner graph by incorporating nodes and edges representing logical observables, which are also defined as products of Pauli operators. We then embed all syndrome data into the extended Tanner graph $\mathcal{G}(V_{\text{data}},V_{\text{check}},V_{\text{logical}},E_{\text{stabilizer}},E_{\text{logical}})$, treating each time as a slice of a dynamic temporal graph. This formulation transforms the decoding problem into a temporal graph classification task, seeking a mapping $f$:
\begin{equation}
    f: \mathcal{G}_t\left(\{S_t\}\right) \rightarrow \mathcal{C}, S_t \in \mathbb{F}_2^{n_s}, \mathcal{C} \in \mathbb{F}_2^k
\end{equation}
where $\{S_t\}$ represents the syndrome sequence, and $\mathcal{C}$ is a binary vector indicating logical qubit flips, with $n_s$ being the number of syndromes and $k$ the number of encoded logical qubits. This binary nature of inputs and outputs reflects the discrete measurement outcomes in quantum experiments.

The graph formulation's expressiveness distinguishes it from existing neural decoders constrained by topological rigidity. Conventional neural approaches, such as Convolution Neural Networks (CNNs) tailored for surface codes, require redesigning network architectures when applied to codes spanning different geometries (e.g., color codes) or non-local structures (e.g., QLDPC codes). In contrast, our extended Tanner graph $\mathcal{G}$ offers a topology-agnostic interface: its nodes and edges dynamically adapt to any stabilizer code's connectivity, whether geometric or algebraic. In addition, the simplification from the common decoding graph allows the neural network to adapt to any experiment settings, and unrestricted by the inaccurate prior noise modeling.

\subsection*{Universal decoding across quantum code families}

To rigorously assess the performance of our decoder, we conduct quantum memory experiments using a variety of quantum error correction codes, encompassing three major code families: color codes with code distances ranging from d=3 to d=11, BB codes represented by two distinct configurations (\qcode{72}{12}{6} and \qcode{144}{12}{12}), and the surface code. For the color codes and BB codes, we utilize simulated data generated under a uniform depolarizing noise model. In contrast, the surface code experiments rely on real experimental data collected from Google's Sycamore quantum device \cite{sycamore2022data}. Critically, \textit{GraphQEC} uses identical neural network architectures for all codes, with adaptations limited to code-specific graph connectivity and hyperparameter optimization—highlighting its versatility.

Table.~\ref{tab:full_compare} summarizes the primary results of the experiments, that GraphQEC consistently outperforms state-of-the-art decoders tailored to individual code families: BP-OSD\cite{roffe_decoding_2020,Roffe_LDPC_Python_tools_2022} for BB codes, Concat-Matching\cite{lee2024color} for color codes, and AlphaQubit\cite{bausch2024learning} for surface codes, as well as the widely used pyMatching\cite{higgott2023sparse} decoder for surface codes. For BB codes and color codes, the experiments are conducted at a fixed physical error rate of p=0.005, reflecting realistic near-term quantum computing conditions \cite{ai2024quantum, evered2023high}. For the surface code, the experimental noise profiles from the Sycamore processor are used. The best decoder for each code are highlighted by bold font. Notably, across all codes, GraphQEC consistently outperforms the traditional baselines, achieving up to 94.6\% lower logical error rates. Further, GraphQEC matches the performance of AlphaQubit, a surface code optimized machine learning decoder despite its own code-agnostic design.

For the simulated experiments (BB and color codes), we further analyze the decoder performance across a range of physical error rates. The results are presented in Fig.~\ref{fig:CompareDecoderHeatmap}. The baseline decoders are initialized with the exact noise model, while our neural network-based decoder is trained exclusively under a physical error rate of $p=0.005$, without adapting to different test conditions. Panel A and B shows the logical error rate (LER), where LER smaller than $10^{-6}$ are not displayed with exact values due to the insufficient number of negative samples required to achieve statistically meaningful results, even after searching tens of millions of samples. Panel C provide with a direct comparison between the GraphQEC decoder and the baseline decoders. The results show that the neural network's advantage becomes more pronounced for more complex codes and lower physical error rates, achieving a suppression rate exceeding 99\% for the \qcode{144}{12}{12} BB code. 

We also indicate the break-even regions in the figure with white borders. Break-even means the logical error rate after decoding is lower than the physical error rate of the encoded qubits, that is the minimum requirement for useful quantum error correction. Our decoder reduces the physical error rate threshold for achieving break-even by $0.1\%$ to $0.2\%$ across different cases (horizon shift), or equivalently reduces the required number of physical qubits by approximately a half (vertical shift).

As our decoder can generalize to different physical error rates, we further calculate the decoding threshold for each code family by fitting the sub-threhold scaling equation respectively (see methods). The deocding threshold is the max allowed physical error rate where we can reduce the logical error rate by increasing the number of encoded qubits, and is a useful metric to evaluate the error correction ability of a code family after decoding. Our result shows that the neural decoder enhances the performance of the color code, improving its parameters from $p_\text{th} = 0.005299$ to $p_\text{th} = 0.007101$. Similarly, for the BB code, the threshold improve from $p_\text{th} = 0.007582$ to $p_\text{th} = 0.007619$. The significantly higher thresholds of color code indicate consistent enhancements across all code sizes, and the less improvement on BB code indicates, given that its logical error rate has been decreased a lot, that the \qcode{144}{12}{12} code benefits more from GraphQEC than the \qcode{72}{12}{6} code.

\subsection*{Optimized designs for efficient decoding}

Our previous sections established the theoretical framework for universal neural decoders and verified their high decoding accuracy, yet significant challenges persist in optimizing efficiency. Real-time decoding, processing syndromes from a quantum computer with constant latency, requires linear time complexity with respect to the number of syndrome measurement cycles, with additional latency per syndrome cycle remaining below the clock frequency of the syndrome measurement.

To address the challenge, we propose a three-phase neural decoding architecture:
\begin{enumerate}
    \item \textit{Spatial Feature Extraction (E-phase):} Embedded multiplicative message passing (MMP) operators process syndrome patterns on Tanner graph slices, explicitly leveraging parity-check constraints.
    \item \textit{Temporal Context Integration (D-phase):} Historical syndrome correlations are captured through linear attention mechanisms, preserving linear inference complexity ($\mathcal{O}(T)$) while enabling parallel training.
    \item \textit{Logical Error Prediction (R-phase):} Final decoding decisions are generated via graph-based pooling, supporting multiple-qubit codes.
\end{enumerate}

\begin{figure}
\centering
\includegraphics[width=\textwidth]{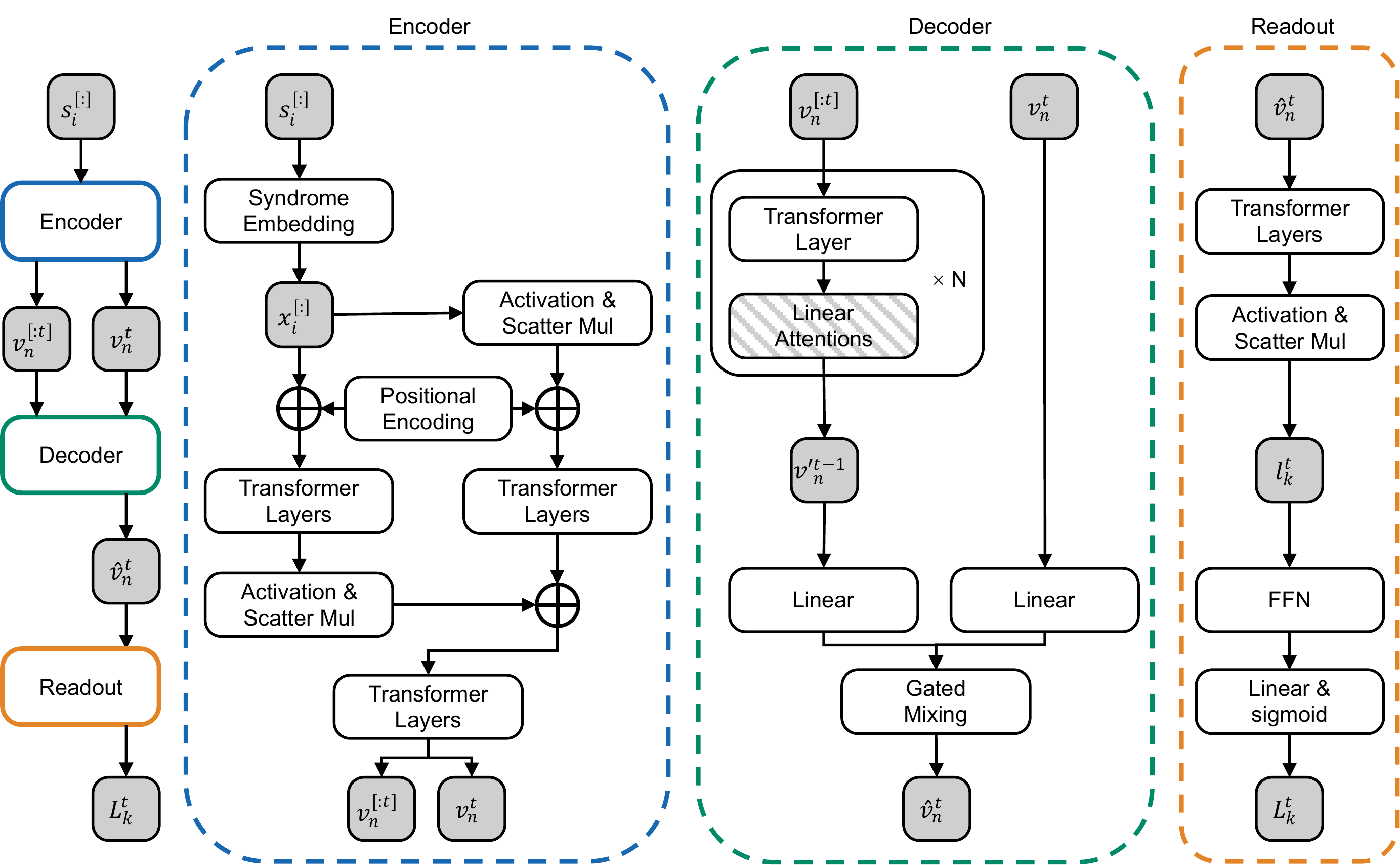}
\caption{\textbf{Architecture of the GraphQEC decoder.} The leftmost flowchart illustrates the high-level structure of our proposed quantum error correction neural network, with colored blocks detailed on the right. White blocks represent operators and gray blocks represent variables. Superscript indices denote time slices, while subscript indices differentiate between node types in the quantum code graph: $i$ for check nodes, $n$ for data nodes, and $k$ for logical nodes. The shadowed linear attention block operates across the temporal dimension, enabling information flow between time slices, while all other operators process spatial relationships within the graph at each individual time step. }\label{fig:ModelStructure}
\end{figure}

The overview of our 3-phase decoder is shown in Fig.~\ref{fig:ModelStructure}, where the encoder (E-phase), decoder (D-phase), and readout (R-phase) modules work in concert to transform syndrome measurements into correction operations. While normalization layers and residual connections are essential components of our implementation (available in our open-source repository), we omit these details in the figure for clarity.

Computational complexity constraints fundamentally shape our architecture design. A key design choice is the use of a standalone D-phase to process temporal correlations of syndromes, separating it from the time-independent E-phase. The time-independent nature of E-phase ensures it to have a linear time complexity with respect to the number of error correction cycles, and even enables parallel computation along the time dimension, resulting in constant decoding latency when sufficient computing resources are available.

Consequently, per-cycle latency requirements only constrain the D-phase, which should scale linearly with the time dimension. While recurrent neural networks (RNNs) naturally satisfy this requirement\cite{sherstinsky2020fundamentals}, their known limitations in long-sequence modeling and training instability make them suboptimal. Instead, we adopt modern linear attention layers—a technique matured through large language model development—which combine parallelizable training with efficient recurrent inference through matrix decomposition techniques \cite{yang2024fla,yang2024gated,yang2024parallelizing}. We also hybridize this approach with full transformer layers to enhance representational capacity, with transformer layers operating on each time step independently to maintain linear time scaling\cite{vaswani2017attention}.

\begin{figure}
\centering
\includegraphics[width=0.66\textwidth]{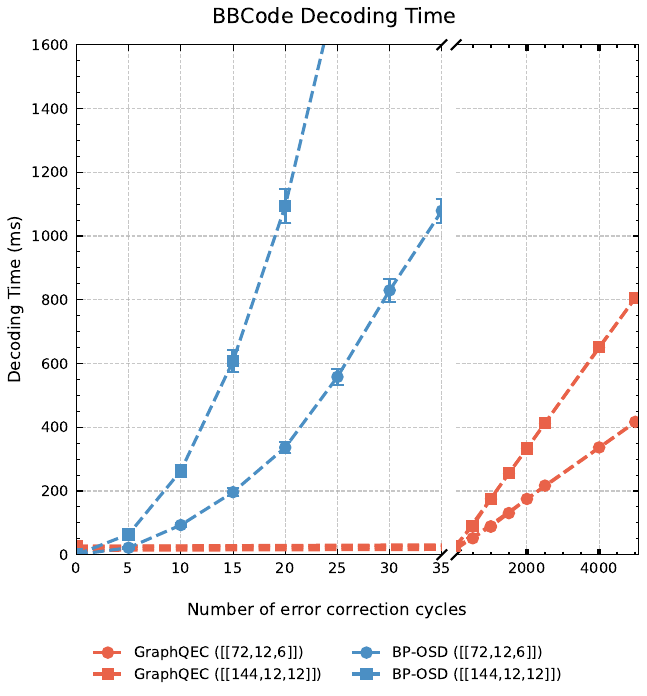}
\caption{
\textbf{Decoding Time for different Decoders on BB code.}
Decoding times are plotted for the d=6 and d=12 BB code. We test it under a physical error rate of $p=0.005$ and take the BP-OSD method for reference. The x-axis is broken to show both short and long decoding times. The GraphQEC decoder is much faster than the BP based decoders and keeps linear time scaling up to thousands of cycles.
}
\label{fig:UniversalDecodingTime}
\end{figure}

To evaluate the efficiency of our proposed architecture, we benchmarked its decoding time against BP-OSD under a physical error rate of $p=0.005$ (Fig.~\ref{fig:UniversalDecodingTime}). Our results demonstrate that the GraphQEC decoder achieves significantly faster performance compared to BP-based methods, maintaining linear time scaling even up to thousands of error correction cycles. This scalability is critical for real-time quantum error correction, as it ensures that decoding latency remains manageable as the number of cycles increases. By contrast, BP-OSD is known to be non-linear, and the decoding time is highly unstable due to the unpredictable unconvergence especially for large number of cycles, as indeicated by the error bar.

These results demonstrate that our neural decoder architecture successfully addresses the efficiency challenges of universal quantum error correction, achieving the linear time complexity and low per-cycle latency required for practical quantum computing applications across a range of code sizes and structures. By combining spatial feature extraction, temporal context integration, and logical error prediction, our decoder provides a scalable and efficient solution for real-time quantum error correction, paving the way for fault-tolerant quantum computing at scale.

\section*{Discussion}

In this work, we have developed a universal framework for quantum error correction decoding using graph neural networks. Our decoder demonstrates superior performance in both accuracy and speed compared to existing approaches. While the framework's architecture is conceptually straightforward, its consistent success across diverse test cases validates the potential of neural network-based approaches.

The challenges in developing neural network decoders differ fundamentally from those of traditional decoders. Rather than algorithmic design, the primary obstacles reside in network training. Our research, alongside other investigations, has demonstrated that the requisite training samples increase exponentially with code scale\cite{Iyer2015StabilizerNP,bausch2024learning}. These challenges illuminate why neural decoders have required substantial time to surpass traditional approaches since their initial conception. Our successful implementation on the \qcode{144}{12}{12} BB code, which establishes a new scale record for neural network decoders, necessitated training with 64 NVIDIA RTX4090 GPUs over a month to achieve its current performance, with training loss diverging across numerous model architectures during our preliminary attempts. The remarkably high decoding accuracy of GraphQEC stems not only from enhanced neural architecture design but also from recent advancements in AI infrastructure.

Despite the substantial training requirements, we contend that neural network approaches represent the most promising path toward real-time decoding for complex quantum error correction codes. While traditional decoders like BP-OSD and maximum weight perfect matching(MWPM) become increasingly slower and inaccurate with both code scale and number of syndrome measurements, neural network decoders encapsulate this complexity in the training phase while maintaining constant inference time. This characteristic is crucial for fault-tolerant quantum devices, where real-time decoding efficiency is paramount and extended pre-deployment training is acceptable.

Several important limitations remain in our current work. Our evaluation focuses solely on quantum memory experiments, which in our temporal graph framework corresponds to static graphs. Many practical quantum operations, such as lattice surgery, involve diverse dynamic graph structures as the Tanner graph evolves. Furthermore, the question of model generalization across different graphs remains open - a capability that could significantly reduce training costs for larger codes.

As a graph-based machine learning approach, our decoder's key strength lies in its direct operation on the mathematical structure of quantum codes. This enables a unique combination of universality and efficiency - the decoder automatically learns optimal strategies from any stabilizer code's graph representation while maintaining constant inference time regardless of code complexity. Our success with large QLDPC codes like the \qcode{144}{12}{12} BB code, which has long challenged traditional decoders, demonstrates how modern machine learning can overcome established bottlenecks in quantum error correction. The decoder's ability to significantly outperform specialized algorithms across diverse code families, while requiring only a single training framework, points to a fundamental shift in how we approach quantum error correction. While the field continues to develop novel quantum codes and hardware architectures, our results suggest that graph neural networks could serve as a universal computational layer bridging theoretical code design and practical quantum computing implementations.

\newpage


\begin{figure}
\centering
\includegraphics[width=\textwidth]{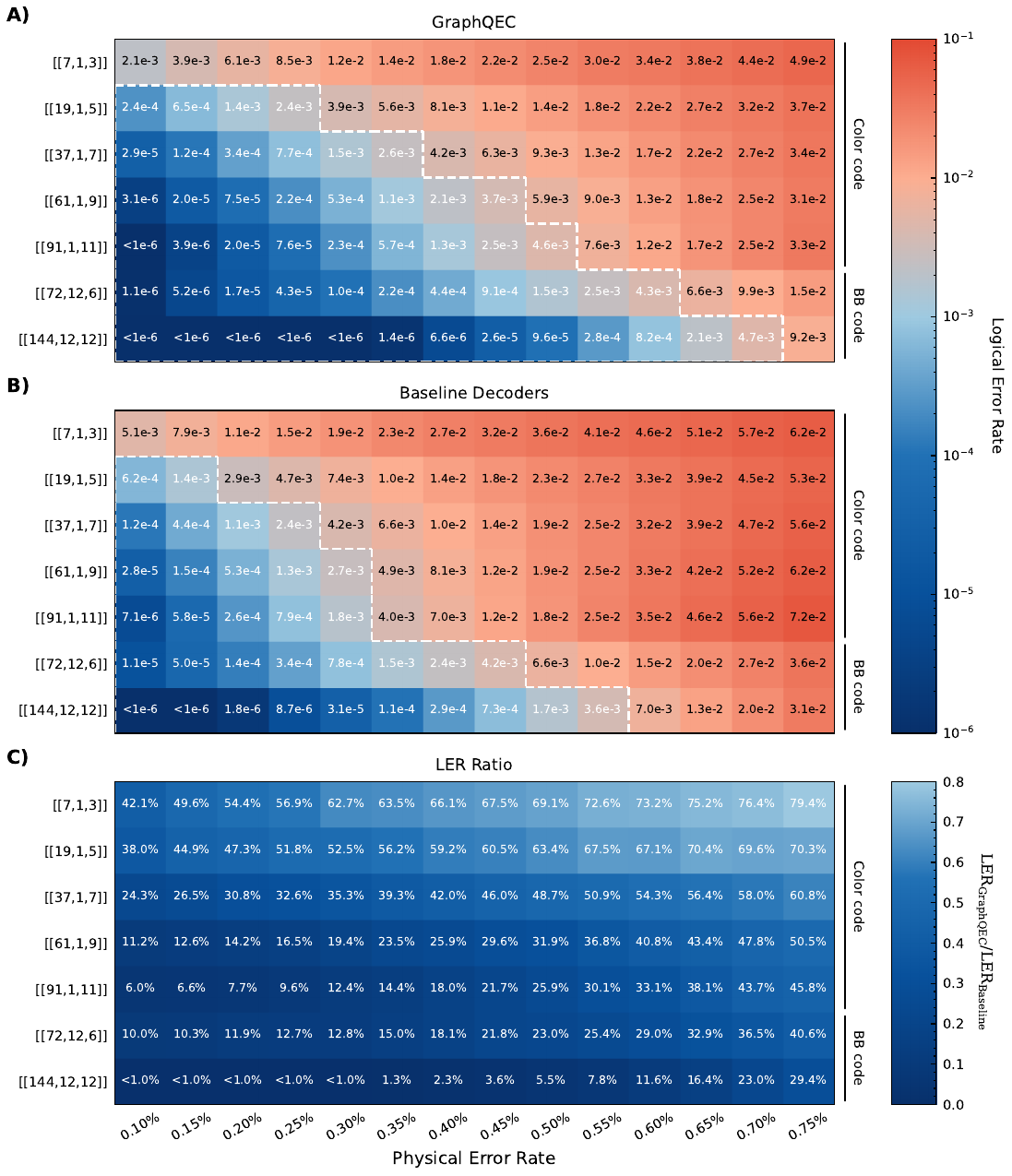}
\caption{
\textbf{Comparative Performance Analysis of Quantum Error Correction Decoders.}
(A) GraphQEC logical error rates (LER) on a logarithmic scale (color bar). 
(B) Baseline decoder (BP-OSD/Concat-Matching) performance under identical conditions. 
White contours indicate break-even regions where logical error rates fall below physical rates. 
(C) Performance ratio \textit{LER\textsubscript{GraphQEC}/LER\textsubscript{Baseline}} demonstrates GraphQEC's superior correction capability, 
with ratios approaching $<1\%$ in low-error regimes.
}
\label{fig:CompareDecoderHeatmap}
\end{figure}

\begin{landscape}
\begin{table}[]
    \centering
    \footnotesize
    \caption{\textbf{Comparison of logical error rates for various quantum error correction decoders.} GraphQEC achieves up to 94.6\% lower logical error rates than BPOSD and other decoders (Bivariate Bicycle, color, and surface codes), demonstrating simultaneous improvements in accuracy and versatility. Error values show mean logical error rates with one standard deviation in parentheses. Missing entries (/) denote decoder-code incompatibilities. Data were acquired under uniform noise conditions: physical error rate \(p=0.005\) for Bivariate Bicycle and color codes, and experimental noise profiles from Sycamore processor for surface codes.}
    \begin{tabular}{ll|lllll}
    \hline
    code & profile & GraphQEC & AlphaQubit & BP-OSD & Concat-Matching & PyMatching \\
    \hline
    BB Code & [[72,12,6]] & $\mathbf{(1.53 \pm 0.02) \times 10^{-3}}$ & / & $(6.65 \pm 0.04) \times 10^{-3}$ & / & / \\
    BB Code & [[144,12,12]] & $\mathbf{(9.55 \pm 0.13) \times 10^{-5}}$ & / & $(1.74 \pm 0.01) \times 10^{-3}$ & / & / \\
    Color Code & [[7,1,3]] & $\mathbf{(2.52 \pm 0.03) \times 10^{-2}}$ & / & $(3.84 \pm 0.04) \times 10^{-2}$ & $(3.64 \pm 0.04) \times 10^{-2}$ & / \\
    Color Code & [[19,1,5]] & $\mathbf{(1.43 \pm 0.02) \times 10^{-2}}$ & / & $(2.36 \pm 0.03) \times 10^{-2}$ & $(2.26 \pm 0.03) \times 10^{-2}$ & / \\
    Color Code & [[37,1,7]] & $\mathbf{(9.35 \pm 0.11) \times 10^{-3}}$ & / & $(1.92 \pm 0.02) \times 10^{-2}$ & $(1.92 \pm 0.02) \times 10^{-2}$ & / \\
    Color Code & [[61,1,9]] & $\mathbf{(5.91 \pm 0.07) \times 10^{-3}}$ & / & $(1.68 \pm 0.02) \times 10^{-2}$ & $(1.85 \pm 0.02) \times 10^{-2}$ & / \\
    Color Code & [[91,1,11]] & $\mathbf{(4.64 \pm 0.05) \times 10^{-3}}$ & / & $(1.58 \pm 0.02) \times 10^{-2}$ & $(1.79 \pm 0.02) \times 10^{-2}$ & / \\
    Surface Code & [[9,1,3]] & $\mathbf{(2.92 \pm 0.01) \times 10^{-2}}$ & $\mathbf{(2.93 \pm 0.02) \times 10^{-2}}$ & $(3.55 \pm 0.01) \times 10^{-2}$ & / & $(4.00 \pm 0.01) \times 10^{-2}$ \\
    Surface Code & [[25,1,5]] & $\mathbf{(2.83 \pm 0.02) \times 10^{-2}}$ & $\mathbf{(2.83 \pm 0.01) \times 10^{-2}}$ & $(4.32 \pm 0.03) \times 10^{-2}$ & / & $(4.43 \pm 0.03) \times 10^{-2}$ \\
    \hline
    \end{tabular}
    \label{tab:full_compare}
\end{table}

\end{landscape}


\clearpage 

%
\bibliography{main} 
\bibliographystyle{sciencemag}

%
%
%
%
%
%


\section*{Acknowledgments}

We thanks Yi-Ming Zhang for his helpful suggestions on quantum circuit simulation, and we would like to acknowledge the use of several open-source projects that were instrumental in this work. We extend our gratitude to the authors of these projects for their valuable contributions.

\paragraph*{Funding:}
This work is partially supported by the National Key R\&D Program of China (NO.2022ZD0160101), the Shanghai Municipal Science and Technology Commission (Grant No. 24DP2600300), and Shanghai Artificial Intelligence Laboratory. 

\paragraph*{Author contributions:}
G. H. developed the algorithms and executed all experiments. G. H. and C. L. performed data analysis. C. L. provided critical guidance in model training and optimization. H.-S. Z. and C.-Y. L. conceived the research. H.-S. Z., C. L., and W. O. supervised the research. All authors collaboratively participated in manuscript drafting, critical revisions, and final approval.

\paragraph*{Data and materials availability:}

The original test results are attached to the supplementary materials, and the corresponding code is open-sourced at https://github.com/Fadelis98/graphqec-paper (Do not ingnore '-' in the link).

\subsection*{Supplementary materials}
Materials and Methods\\
Supplementary Text\\
Figs. S1 to S4\\
Tables S1 to S2\\
References \textit{(41-\arabic{enumiv})}\\ 
Data S1 to S2


\newpage


\renewcommand{\thefigure}{S\arabic{figure}}
\renewcommand{\thetable}{S\arabic{table}}
\renewcommand{\theequation}{S\arabic{equation}}
\renewcommand{\thepage}{S\arabic{page}}
\setcounter{figure}{0}
\setcounter{table}{0}
\setcounter{equation}{0}
\setcounter{page}{1} 


\begin{center}
\section*{Supplementary Materials for\\ \scititle}


Gengyuan Hu$^\text{1}$,
Wanli Ouyang$^{\text{1,2}\ast}$,
Chao-Yang Lu$^\text{3,4,5,6}$,
Chen Lin$^{7\ast}$,
Han-Sen Zhong$^{\text{1,8}\ast}$ \\

\small$^\ast$Corresponding author. Email: wlouyang@ie.cuhk.edu.hk \\
\small$^\ast$Corresponding author. Email: chen.lin@eng.ox.ac.uk \\
\small$^\ast$Corresponding author. Email: zhonghansen@pjlab.org.cn \\

\end{center}

\subsubsection*{This PDF file includes:}
Materials and Methods\\
Supplementary Text\\
Figures S1 to S4\\
Tables S1 to S2\\
Captions for Data S1 to S2

\subsubsection*{Other Supplementary Materials for this manuscript:}

Data S1 to S2

\newpage


\subsection*{Materials and Methods}




\subsubsection*{Neural network design}

\paragraph{Encoder Phase}

In the encoding phase, we leverage the bipartite structure of the Tanner graph to process information in both the check-node and data-node feature domains. The input syndromes $s_i^{[:]} \in \mathbb{F}^2$ are embedded as initial hidden features $x_i^{[:]} \in \mathbb{R}^h$ on the check nodes, where $h$ is the hidden feature dimension. The initial data-node features $v_n^{[:]} \in \mathbb{R}^h$ are created through a multiplicative message-passing mechanism, defined as:
\begin{equation}
v_n^{[:]} = \prod_{j \in \mathcal{N}(n)} \tanh(x_j^{[:]}),
\end{equation}
where $\mathcal{N}(n)$ denotes the set of neighboring check nodes connected to the data node $n$, and $\sigma(\cdot)$ is an activation function (e.g., $\tanh$). This operation aggregates the check-node features to the corresponding data nodes based on the bipartite graph structure.

To refine the node features, we employ self-attention transformer layers independently for both check nodes and data nodes. Specifically, the check-node features are updated as:
\begin{equation}
x_{\text{out}} = \text{Transformer}(x^{[:]} + \text{PE}_{\text{check}}),
\end{equation}
where $\text{PE}_{\text{check}}$ is a learnable positional encoding for the check nodes. Similarly, the data-node features are updated as:
\begin{equation}
v_{\text{out}} = \text{Transformer}(v^{[:]} + \text{PE}_{\text{data}}),
\end{equation}
where $\text{PE}_{\text{data}}$ is the learnable positional encoding for the data nodes. The transformer layers used in this module are fully connected transformer layers, which are implemented as a post-norm variant of the standard multi-head self-attention mechanism \cite{vaswani2017attention, xiong2020layer}. 

After updating the features independently, the check-node features are aggregated to the data nodes using the same multiplicative scattering mechanism:
\begin{equation}
\text{out} = v_{\text{out}} + \prod_{j \in \mathcal{N}(n)} x_{\text{out}, j}.
\end{equation}
The aggregated features are then passed through additional transformer layers to fuse the information from both domains:
\begin{equation}
\text{out}_{\text{final}} = \text{Transformer}(\text{out}).
\end{equation}

Finally, the output features are projected to the desired dimension using a normalization layer followed by a linear transformation:
\begin{equation}
\text{output} = \text{Linear}(\text{Norm}(\text{out}_{\text{final}})),
\end{equation}
where $\text{Norm}$ is the root-mean-square normalization (RMSNorm) \cite{zhang2019root}, which stabilizes the training process by normalizing the features across the hidden dimension.

This bipartite-attention mechanism enables the encoder to process information in both the data and check feature domains, with the multiplicative aggregation ensuring isomorphism with the parity-check constraints. Additionally, the use of learnable positional encodings and transformer layers allows the model to capture complex dependencies within each domain. To further enhance performance, we optionally employ regional compilation and optimize the transformer layers for reduced overhead.

\paragraph{Decoder Phase}

In the decoding phase, the goal is to refine the encoded features and generate the final feature that related to the logical error probabilities for each data node. The decoder operates on the sequence of encoded states, which are the outputs of the encoding phase. These states are processed through a hybrid gated delta network architecture, which combines the GatedDeltaNet linear attention with traditional multi-head attention mechanism to capture both temporal and structural dependencies\cite{yang2024gated,yang2024parallelizing}.

The outputs of the linear attention and transformer layers are alternately combined to fuse temporal and structural information. Specifically, the cycle states are updated as:
\begin{equation}
v^{(l)} = \text{Transformer}(\text{GatedDeltaNet}(v^{(l-1)})),
\end{equation}
where $v^{(l)}$ denotes the hidden states at the $l$-th layer of the decoder. This alternating update mechanism ensures that the model effectively integrates both temporal and graph-based dependencies. During transformer layers, the time dimension is treated as a batch dimension, and vice versa during GatedDeltaNet layers. This approach, combined with linear attention, significantly reduces the computational complexity from $O(n^2L^2)$ for a fully connected causal transformer to $O(n^2 + nL)$, where $n$ is the number of nodes and $L$ is the number of time steps.

The syndromes in the last cycle are treated differently for better training efficiency. An independent gated recurrent unit is employed to mix the features from the final cycle with the previous cycle's features. The mixing operation is defined as:
\begin{align}
\hat{v}_n^t &= \sigma v_n^{t-1} + (1-\sigma)v_n^t, \\
\sigma &= \tanh(W_1 v_n^{t-1} + W_2 v_n^t + b),
\end{align}
where $W_1$, $W_2$, and $b$ are learnable parameters, and $v_n^t$ represents the hidden features of node $n$ at time step $t$.

\paragraph{Readout Phase}

In the readout phase, the neural module extract the information in the data node features, then aggregate them into the logical nodes and explain them as the probability of wrong readout result on each logical qubit. Again we use the multiplicative message passing here to aggregate node features to model the parity check property:
\begin{equation}
l_k^{[:]} = \prod_{j \in \mathcal{N}(k)} \tanh\left(\text{Transformer}(v_j^{[:]})\right),
\end{equation}

\subsubsection*{Quantum Memory Experiment}

\paragraph*{Experiment setting}

We assess the efficacy of quantum error correction codes through quantum memory experiments, which serve as a standardized framework for evaluating code performance. In these experiments, the physical qubits undergo a sequence of repeated syndrome measurements, which functions as logical identity operations. The experimental protocol consists of $d$ consecutive syndrome extraction cycles, where $d$ represents the code distance. This specific choice of cycle count, while a convention rather than a fundamental requirement, establishes a meaningful benchmark that correlates with the code's theoretical error-correction capabilities. Each syndrome extraction cycle comprises a series of quantum operations including state preparation, gate operations, and measurements, all of which are subject to noise in realistic implementations.

Following the completion of these $d$ cycles, we perform measurements on the data qubits to extract the final state information. The measurement outcomes, along with the syndrome measurement results collected throughout the cycles, are processed using a decoding algorithm to determine the most likely error that occurred during the experiment. Since we know the initial state of the system, we can determine the actual errors from the readout outcomes, enabling us to evaluate the decoder's performance.

\paragraph*{Noise simulation}
As the circuit of quantum memory experiment is Clifford, it can be efficiently simulated by classical algorithms\cite{bravyi2016improved}. We simulate the experiment on a classical computer using the STIM\cite{gidney2021stim} package. To model realistic quantum hardware behavior, we employ a circuit-level depolarizing model that introduces stochastic errors into the quantum circuit operations.

The depolarizing noise model is parameterized by four error rates. During periods between gate operations, each qubit undergoes depolarizing noise with probability $p_\text{idle}$. After each quantum gate operation, the qubits involved in the gate experience depolarizing noise with probability $p_\text{gate}$. Following each qubit reset operation, the qubit may be flipped to the opposite state with probability $p_\text{reset}$. During the readout process, each measurement outcome is flipped with probability $p_\text{meas}$. The depolarizing noise here refers to random Pauli errors: for single-qubit operations, it corresponds to probability $p/3$ for each $X$, $Y$, and $Z$ pauli error, while for two-qubit operations such as CNOT, it corresponds to probability $p/15$ for each non-identity Pauli error on the control and target qubits. A canonical choice of $p_\text{idle} = p_\text{gate} = p_\text{meas} \equiv p$ results in the uniform depolarizing noise model, which is widely used in quantum error correction studies.

\paragraph*{Metrics}

The primary performance metric in our analysis is the logical error rate after decoding. Formally, given the logical measurement outcome $\hat{L}$ and decoder prediction $L$, we identify an error when $L_k \neq \hat{L}_k$ for each logical qubit $k$. As errors accumulate through successive syndrome measurement cycles, we characterize the accumulated logical error rate $p_l$ as a function of the cycle count $r$. Under the assumption of independent and identically distributed errors per cycle, we use the ansatz:
\begin{equation}\label{eq:LER}
    p_l(r) = \frac{1}{2}(1-\left(1 - 2p_c\right)^r)
\end{equation}
where $p_c$(or LER) represents the per-cycle logical error rate, which serves as a fundamental metric for quantifying the performance of a quantum error correction code. One can verify that $p_l(0) = 0$, $p_l(1) = p_c$, and $p_l(\infty)=\frac{1}{2}$. We can further define the fidelity $F(r) = 1 - 2p_l(r)$, which leads to an expression describing the exponential decay of fidelity:
\begin{equation}
    F(r) = (1 - 2p_c)^r
\end{equation}
This ansatz aligns with previous work on neural network decoders \cite{google2021exponential,bausch2024learning}, but differs slightly from the formulation more commonly used in quantum LDPC decoder research \cite{bravyi2024high,panteleev2021degenerate,gong2024toward,wolanski2024ambiguity}, where researchers typically assume an exponential decay in the probability of avoiding any logical error.

It is important to note that Equation~\eqref{eq:LER} assumes perfect readout operations, such that $p_l=0$ when no syndrome measurement cycles are performed ($r=0$). In our numerical simulations, we implement perfect measurements, allowing direct calculation of $p_c$ through the inverse of Equation~\eqref{eq:LER}:
\begin{equation}
    p_c = \frac{1}{2}\left(1-F(r)^{\frac{1}{r}}\right)
\end{equation}
However, for real experimental data where measurement errors are unavoidable, we employ linear regression to separate the contribution of measurement errors from the cycle-dependent error accumulation:
\begin{equation}
    \log F(r) = \log F(0) + r \log(1-2p_c)
\end{equation}
where the intercept $\log F(0)$ is expected to be close to 0, as perfect fidelity would yield $F(0)=1$.

\paragraph{Sub-threshold Scaling}

The decoding threshold is a key property that evaluates the error correction ability of a code family. For each code family, the decoding threshold is the phase transition point related to the physical error rate, where increasing the number of encoded qubit reduce the logical error rate only if the physical error rate is below the decoding threshold. Good decoders will improve the decoding threshold for a given code family, so that it is widely used as a metric for decoders as well.

\begin{figure}
\centering
\includegraphics[width=\textwidth]{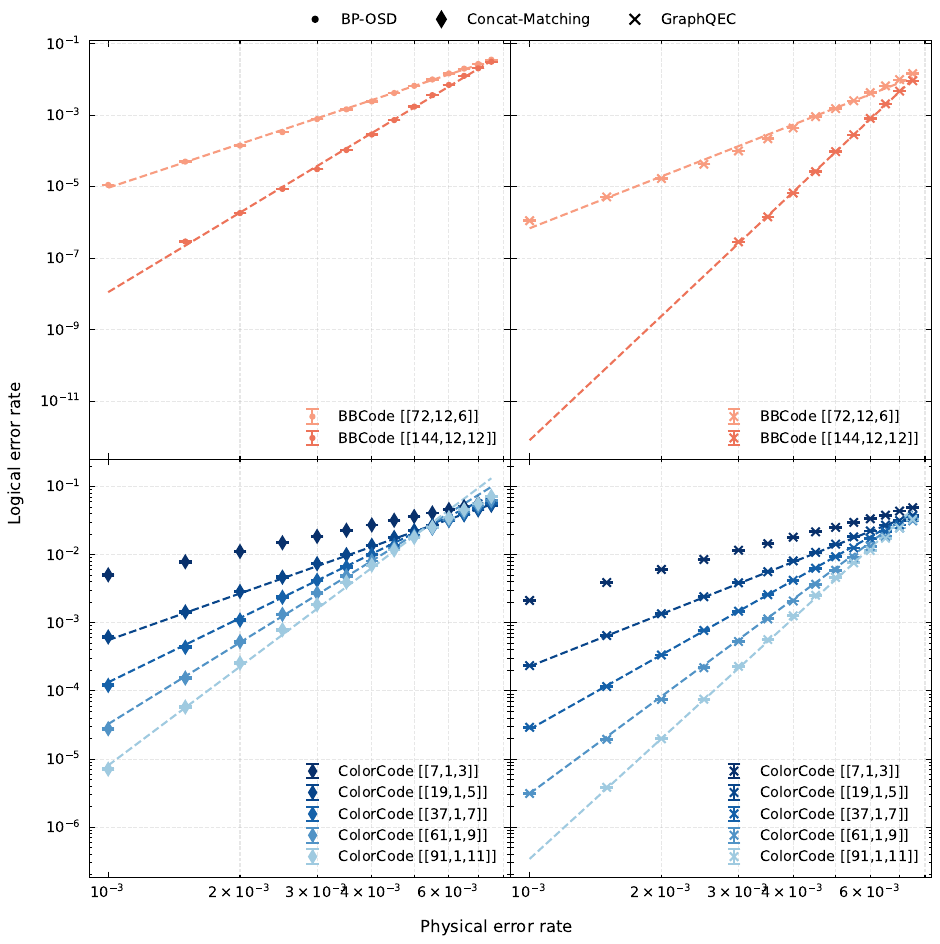}
\caption{\textbf{The sub-threshold scaling of decoder performance on different code.} The data points with different marker represents different decoders. The dash-lines is the fitting curve, where fitting parameters are decided separately for each decoder and each code family, and is shared inside each group. The $d=3$ color code is excluded from the fitting to reduce the influence of finite size effect.}\label{fig:SubThreshold}
\end{figure}

Figure~\ref{fig:SubThreshold} presents the complete test results used to calculate the decoding threshold values. The data displayed here is identical to that shown in Fig.~\ref{fig:CompareDecoderHeatmap} in the main text; however, those figures imposed a lower limit on the logical error rate (LER) at $10^{-6}$. In contrast, we present the unclipped values here, including cases where the LER is too small to be reliably estimated. 

We estimate the decoding performance by fitting the sub-threshold behavior using the established scaling formula~\cite{Wang2003confinement,xu2024constant}:

\begin{equation}
    \text{LER} = A(p/p_{\text{th}})^{\alpha n^\beta / 2},
\end{equation}\label{eq:sub_threshold}

where $p$ denotes the physical error rate, $n$ represents the number of physical qubits required in a minimal encoding unit, and LER is the logical error rate. Among the fitting parameters, two are particularly noteworthy: the threshold physical error rate $p_\text{th}$ and the scaling factor $\beta$. The threshold $p_\text{th}$ signifies the maximum physical error rate at which increasing the code distance effectively reduces the logical error rate. Meanwhile, $\beta$ quantifies the efficacy of increasing the block length in suppressing errors.

\subsubsection*{The training pipeline}

\paragraph*{Parallel Training}\label{sec:parallel_training}

Training efficiency is a critical factor when scaling up neural network decoders, as the volume of required training data grows significantly with the number of stabilizer generators. To address this challenge, we approach the problem from both data generation and neural network design perspectives.

\begin{figure}
\centering
\includegraphics[width=\textwidth]{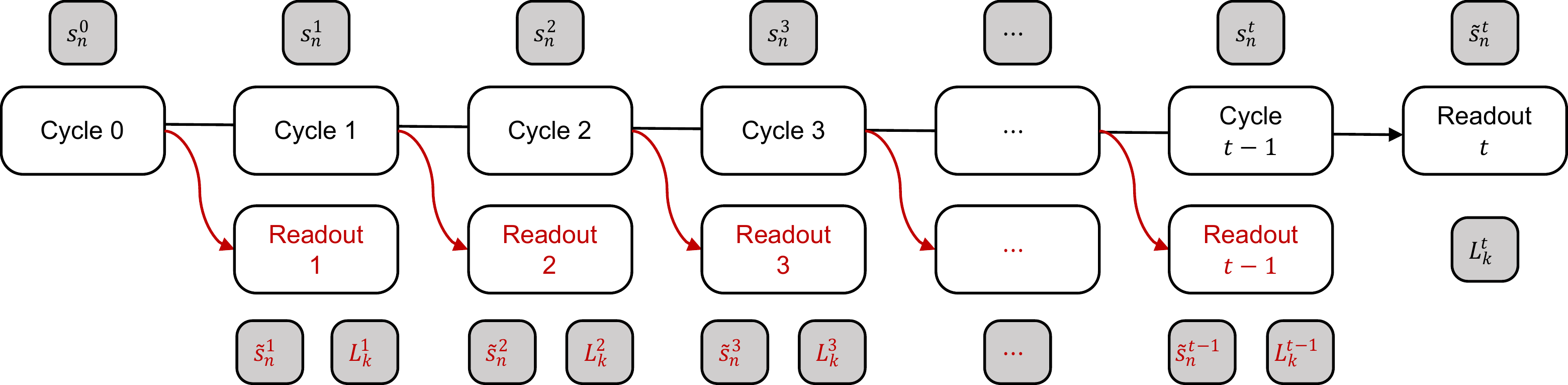}
\caption{\textbf{Construction of training data.} Beside the syndrome sequence of a normal quantum circuit, we extend the circuit with an extra branch (red ones) at each cycle to generate extra pseudo syndromes for parallel training.
}\label{fig:IncrementalSampling}
\end{figure}

In traditional simulations, logical observables are measured only at the end of the quantum circuit, providing a single label for an entire syndrome sequence. This results in sparse supervision signals during training. To mitigate this limitation, we adopt an incremental simulation strategy that generates additional pseudo syndromes and corresponding labels at each time step. This is achieved by introducing extra branches to the quantum circuit at the beginning of each cycle, as illustrated in Fig.~\ref{fig:IncrementalSampling}.

Our framework employs two distinct types of circuits for error correction cycles: cycle circuits and readout circuits. Cycle circuits perform Pauli measurements on check qubits, while readout circuits directly measure data qubits and compute stabilizer results as classical functions of these measurements. This approach is feasible only in simulated environments, as measuring data qubits in real quantum systems would collapse the logical state, preventing further syndrome measurements.

\begin{figure}
\centering
\includegraphics[width=\textwidth]{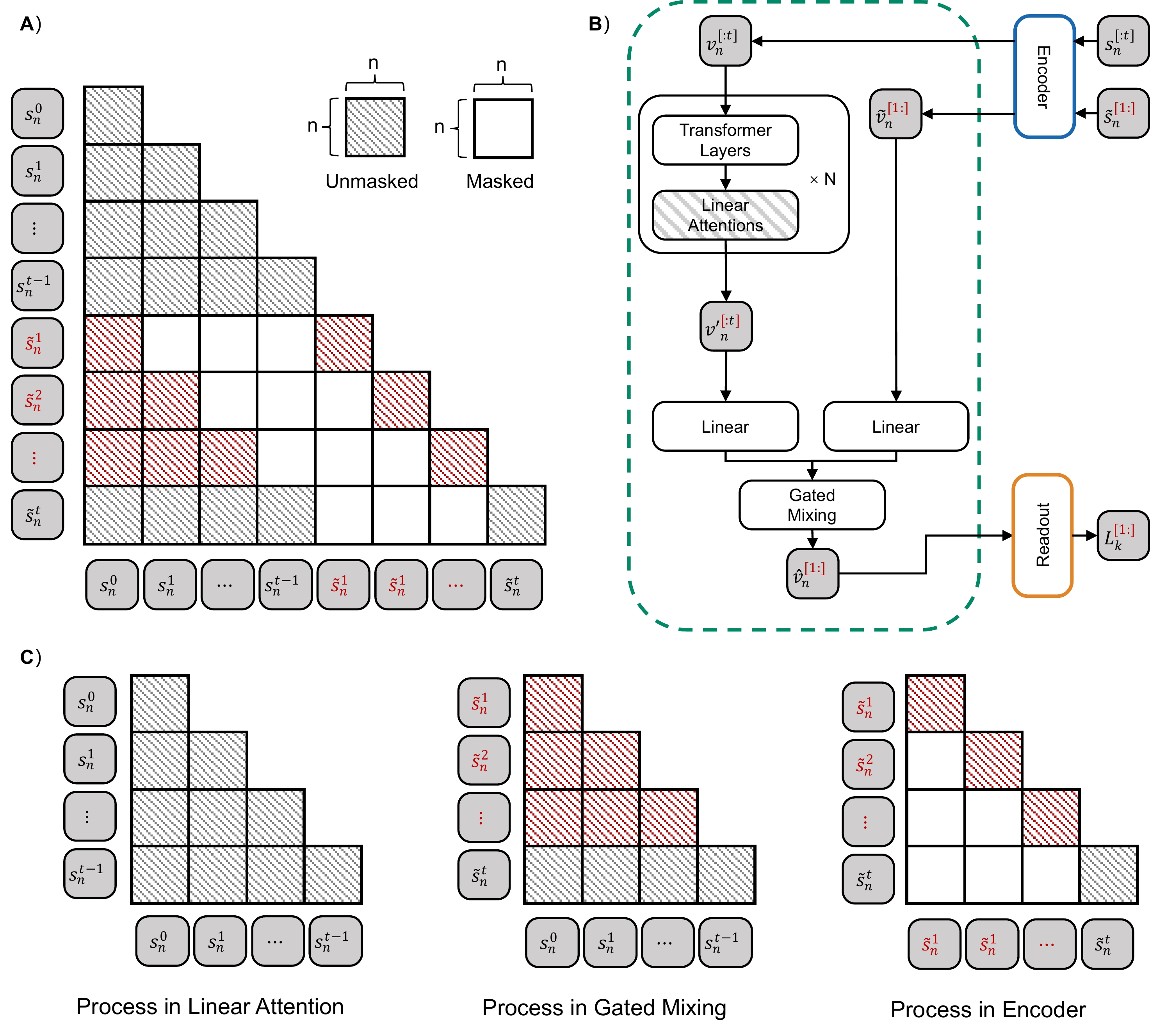}
\caption{\textbf{Parallel training with pseudo syndromes and labels.} Red symbols and blocks indicate information that appears only during the training stage. A) The effective attention mask of the decoder module. The matrix has a block structure where each block represents all nodes at a single time slice. When extra pseudo syndromes are present (shown in red), the mask matrix becomes sparse. B) Variable flow in the parallel training process, where some variables contain an additional time dimension compared to the original case. C) The attention mask matrix can be decomposed into three dense components, enabling efficient non-sparse computation during training.
}\label{fig:IncrementalTraining}
\end{figure}

The architecture of our neural network is tailored to this data structure. As shown in Fig.~\ref{fig:IncrementalTraining}, we designed the network to be hardware-efficient by enabling parallelization along the time dimension. Fig.~\ref{fig:IncrementalTraining}(A) depicts the structure of the effective attention mask matrix during training. The attention mask matrix, derived from the self-attention mechanism \cite{vaswani2017attention}, captures relationships within a sequence by assigning weights that determine which parts of the input should be prioritized or ignored during computation. A causal attention mask ensures that later hidden states are updated based only on earlier signals, resulting in a lower triangular mask matrix. We use the term "effective attention mask matrix" to describe the perception range of nodes, although it is not directly used in attention computations.

Without pseudo syndrome readouts, the effective attention mask is block-causal: each node can process syndromes from all nodes at the current and previous time steps. If the attention mask is purely causal, attention computations can be performed very efficiently. If the mask is diagonal, computations can be fully parallelized along the time dimension.

With the inclusion of pseudo syndrome readouts, the effective attention mask becomes sparse: nodes in readout circuits can access all previous cycle syndromes but are restricted from accessing other readout syndromes. However, the specific structure of this sparse attention allows us to decompose it into three dense components:
\begin{itemize}
    \item The block-causal cycle-to-cycle component,
    \item The block-causal readout-to-cycle component,
    \item The diagonal readout-to-readout component.
\end{itemize}
This decomposition offers computational advantages, as each dense component can be processed efficiently using optimized algorithms. The cycle-to-cycle component utilizes standard causal attention mechanisms, the readout-to-cycle component leverages similar causal structures, and the diagonal readout-to-readout component can be fully parallelized. This is why we separate readout syndromes in our neural decoder architecture, with each component handled by a dedicated module to maximize computational efficiency.

Another aspect of parallel training is the loss function $\mathcal{L}$. Currently, we compute the total loss as the average binary cross-entropy (BCE) loss across all time steps:
\begin{equation}
    \mathcal{L}_{\text{total}} = \frac{1}{T}\sum_t\mathcal{L}_\text{BCE}.
\end{equation}
However, we acknowledge that noise accumulates over cycles, making later cycles inherently less accurate than earlier ones. While we believe that optimal loss weights should vary across cycles, this consideration has not yet been incorporated and remains an open question for future research.

\paragraph*{Hyper-parameter Tuning}

The proposed model structure encompasses a vast hyper-parameter space. To manage this complexity, we performed grid searches on the $d=3$ color code, $d=5$ color code, and the \qcode{72}{12}{6} BB code to identify suitable ranges for key hyper-parameters. The final model configurations were selected based on these systematic grid search results. Table~\ref{tab:ModelConfigures} summarizes the configurations used in our final experiments. Notably, different codes require varying model sizes to achieve optimal performance. For each code, we also determined the optimal number of training samples, learning rate, and batch size through extensive experimentation. The key hyper-parameters for each decoder reported in this paper are listed in Table~\ref{tab:CodeParams}, with the complete configurations available in the open-sourced project to facilitate reproducibility.

Although we have not exhaustively explored the entire hyper-parameter space, we provide the following empirical guidelines for readers interested in training their own neural network decoders:

\begin{itemize}
    \item For small codes, the encoder module plays a relatively minor role. However, its importance grows significantly for larger codes. Allocating more parameters to the encoder module enhances the neural network's ability to process signals from check nodes effectively.
    \item The decoder module substantially improves the neural network's performance in later cycles, with the number of layers being particularly critical. While increasing the number of decoder layers can boost performance, it also reduces training stability. Wider layers may enhance performance, but at least two decoder layers are necessary for effective operation.
    \item The readout module determines the upper limit of decoding accuracy and accelerates learning during the early stages of training. Typically, neural networks first learn to decode early cycles, a capability primarily enabled by the readout module. Although decoder layers alone can suffice for smaller codes, we recommend employing a deep readout module for more complex codes to achieve better performance.
\end{itemize}

\subsubsection*{Benchmark settings}

\paragraph*{Simulated Experiments}

To evaluate the logical error rate of the decoder, we repeatedly decode random data sampled from simulated quantum circuits. Mathematically, we estimate the probability of decoding errors using the frequency obtained from Monte Carlo experiments. The uncertainty of this estimation decreases as more data is accumulated.

Let $ p_l $ denote the ground-truth logical error rate. The decoding result follows a Bernoulli distribution with parameter $ p_l $, where success (error) occurs with probability $ p_l $, and failure (no error) occurs with probability $ 1 - p_l $. For a random variable $ X $ following this distribution, the variance is given by:
\begin{equation}
    \sigma^2 = \mathbb{E}[X^2] - \mathbb{E}[X]^2 = p_l - p_l^2 = p_l(1 - p_l).
\end{equation}

For $ n $ independent trials, the sample mean $ \hat{p_l} $ serves as an unbiased estimator of $ p_l $, with variance $ p_l(1 - p_l)/n $. We use this formula to quantify the uncertainty of our results in all simulated experiments.

To achieve a desired relative precision $ \epsilon $ in our estimation, we require:
\begin{equation}
    \frac{\sqrt{\text{Var}(\hat{p_l})}}{\mathbb{E}[\hat{p_l}]} = \frac{\sqrt{p_l(1 - p_l)/n}}{p_l} \leq \epsilon.
\end{equation}

Solving for $ n $, we obtain:
\begin{equation}\label{eq:num_sample}
    n \geq \frac{1 - p_l}{\epsilon^2 p_l}.
\end{equation}

However, this formula depends on a prior value of $ p_l $, which is determined by the experiment itself. To circumvent this issue, we instead fix the number of failure samples to control the relative error in the benchmark. Specifically, we require $ n_\text{fail} \geq C $, ensuring that the error $ \epsilon $ is bounded by:
\begin{equation}
    n_\text{fail} \approx \mathbb{E}[n p_l] = \frac{1 - p_l}{\epsilon^2} \geq C.
\end{equation}


\paragraph*{The sycamore experiments}

In the sycamore experiments, we adopt the methodology outlined in prior work \cite{bausch2024learning}, where neural decoders are trained independently on distinct subsets. These subsets are defined by their measurement basis and physical qubit regions. Specifically, we employ a 2-fold cross-validation strategy, splitting the data by index parity within each subset. For each fold of every subset, pretraining data is generated using the XEB error model, and the first 19,880 experimental samples are used to fine-tune the model.

The prior work also evaluated the performance of an ensemble of neural decoders by training 20 models with different random seeds. However, due to computational constraints, we train only a single model for each subset. As shown in Table~\ref{tab:full_compare}, both our results and the baseline are non-ensemble outcomes: GraphQEC represents a single-model result, while AlphaQubit reports the average across 20 models.







\begin{figure}
\centering
\includegraphics[width=0.66\textwidth]{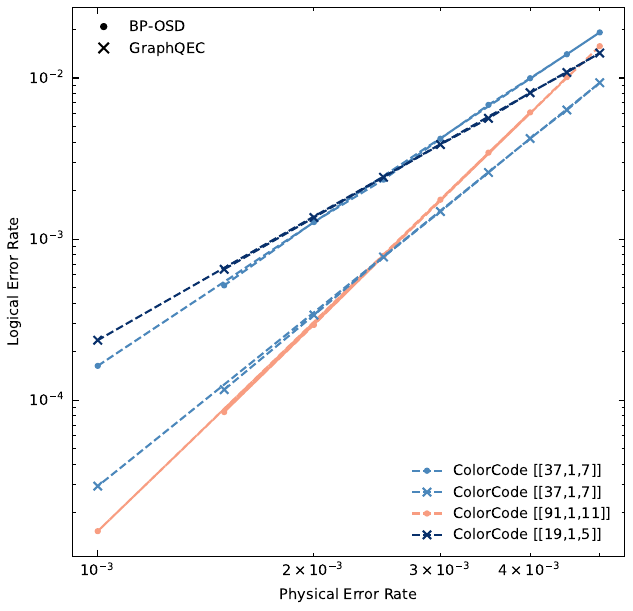}
\caption{\textbf{The logical error rate of selected color codes with different decoders.} By decoding with GraphQEC, it requires about half of physical qubits to achieve comparable logical error rate compare to decoding with BP-OSD.}\label{fig:ReduceOverhead}
\end{figure}

\begin{table}
    \centering
    \caption{\textbf{Comparison of Model Configurations.} We implemented four different configurations of our neural network with parameter counts ranging from 2.6M (Small) to 19.8M (Large). The parameter counts exclude embedding and positional encoding parameters. For mixture-of-experts (MoE) components, parameters are counted only once across all experts.}
    \label{tab:ModelConfigures}
    \begin{tabular}{|l|c|c|c|c|}
        \hline
        \textbf{Config} & \textbf{Small (2.6M)} & \textbf{Medium (8.1M)} & \textbf{Medium+ (10.1M)} & \textbf{Large (19.8M)} \\
        \hline
        \textbf{encoder\_dim} & 96 & 128 & 128 & 192 \\
        \textbf{decoder\_dim} & 128 & 192 & 192 & 256 \\
        \textbf{readout\_dim} & 128 & 128 & 128 & 192 \\
        \textbf{num\_encoder\_layers} & 2 & 4 & 4 & 5 \\
        \textbf{num\_decoder\_layers} & 3 & 3 & 6 & 4 \\
        \textbf{num\_readout\_layers} & 6 & 16 & 16 & 16 \\
        \textbf{num\_heads} & 8 & 8 & 8 & 8 \\
        \textbf{scatter\_activation} & tanh & tanh & tanh & tanh \\
        \textbf{scatter\_fn} & mul & mul & mul & mul \\
        \textbf{ffn\_dim\_multiplier} & 3.0 & 3.0 & 3.0 & 3.0 \\
        \textbf{multiple\_of}& 32 & 32 & 32 & 32\\
        \textbf{norm\_eps} & $1 \times 10^{-5}$ & $1 \times 10^{-5}$ & $1 \times 10^{-5}$ & $1 \times 10^{-5}$ \\
        \hline
    \end{tabular}
\end{table}

\begin{table}
\centering
\caption{\textbf{Summary of key training parameters for each neural decoder.} The model size refers to the configurations listed in \ref{tab:ModelConfigures}, and the batch size of \qcode{9}{1}{11} color code changed from 512 to 1024 at the half of its training process.}
\label{tab:CodeParams}
\begin{tabular}{|c|c|c|c|c|c|c|}
\hline
Code & $[[n,k,d]]$ & Model size & Pretrain steps & Learning rate & Batch size & Pretrain length \\ \hline
BB code & \qcode{72}{12}{6} & Large & 5M & 0.0001 & 512 & 24 \\ 
BB code & \qcode{144}{12}{12} & Large & 10M & 0.0003 & 256 & 24 \\
Color code & \qcode{7}{1}{3} & Small & 800k & 0.0003 & 1024 & 18 \\
Color code & \qcode{19}{1}{5} & Median & 2M & 0.0003 & 1024 & 24 \\
Color code & \qcode{37}{1}{7} & Median & 3M & 0.0003 & 1024 & 24 \\
Color code & \qcode{61}{1}{9} & Large & 8M & 0.0001 & 512 & 24 \\
Color code & \qcode{91}{1}{11} & Large & 10M & 0.0001 & $512^\ast$ & 24 \\
Surface code & \qcode{9}{1}{3} & Median+ & 600k & 0.0005 & 2048 & 25 \\
Surface code & \qcode{25}{1}{5} & Median+ & 1.6M & 0.0005 & 2048 & 25 \\ \hline
\end{tabular}
\end{table}






\end{document}